# The Effect of Hop-count Modification Attack on Random Walk-based SLP Schemes Developed for WSNs: a Study

**Manjula Raja**[a,1], **Anirban Ghosh**[b,2], **Chukkapalli Praveen Kumar**[c,3], **Suleiman Samba** [d,4], **C N Shariff** [e,5]

[1]Department of Computer Science and Engineering, SRM University AP, Guntur District, 522502, Andhra Pradesh, INDIA
[2]Department of Electronics and Communication Engineering, SRM University AP, 522502, Andhra Pradesh, INDIA
[3]Department of Computer Science and Engineering, SRM University AP, 522502, Andhra Pradesh, INDIA
[4]Department of Computer Science and Engineering, SRM University AP, Guntur District, 522502, Andhra Pradesh, INDIA
[5]Department of Artificial Intelligence and Machine Learning, BITM, Ballari-Hosapete Road, 583104

**Abstract** Source location privacy (SLP) has been of great concern in WSNs when deployed for habitat monitoring applications. The issue is taken care of by employing privacy-preserving routing schemes. In the existing works, the attacker is assumed to be *passive* in nature and backtracks to the source of information by eavesdropping the message signals. In this work, we try to understand the impact of *active* attacks by proposing a new hybrid attack model consisting of both active and passive attacks. The proposed model is then applied to three existing TTL-based random walk SLP solutions - phantom routing scheme (PRS), source location privacy using randomized routes (SLP-R), and position independent section-based scheme (PSSLP). The performance of the algorithms in terms of privacy metrics is compared in the case of pure passive attack and hybrid attack of varying intensity. The results indicate a significant degradation in the privacy protection performance of the reference algorithms in the face of the proposed hybrid attack model indicating the importance and relevance of such attacks. It is further observed that the hybrid attack can be optimized to increase the vulnerability of the existing solutions.

.

*Thanks to the title

[a]e-mail: rajamanjula12@gmail.com
[b]e-mail: aniz.ghosh@gmail.com
[c]e-mail: praveenchukkapalli09@gmail.com
[d]e-mail: Sambaj418@gmail.com
[e]e-mail: shariff1@gmail.com

## 1 Introduction

Wireless Sensor Networks (WSN) have evolved quite a lot over the last decade leading to the development of the Internet of Things (IoT). IoT has opened doors to a plethora of applications that can be deployed in the wilderness with remote monitoring, are of a self-healing nature, or can power themselves for sustenance. To cite a few examples military personnel [23], healthcare systems [7, 21], or even endangered species [1, 24] can be equipped with smart devices and monitored remotely. However, such networked deployment makes the assets vulnerable by revealing the location of the soldiers on the battlefield or endangered species in a sanctuary. In this context, the threat can be two-pronged, one in which an attacker eavesdrops on the communication links (*passive attacks*) to backtrack to the origin of the communication or the other in which it can even meddle with the sensor nodes (*active attacks*) to fake biological and chemical attacks, hop count modification attacks, packet dropping attacks etc. [5].

To further explain the two attacks just described let us consider the famous *panda-hunter* game model proposed in [18]. In this model, a large number of sensor nodes are deployed to monitor the pandas. These sensors sense the presence of pandas and report the location to the sink (also known as the base station (BS)). This information is relayed on a hop-by-hop basis due to the limited communication range of the sensor nodes. Meanwhile, the attacker (i.e., the hunter) is all set to poach the animal. In the absence of any location protection technique, the hunter can easily achieve the objective by *passively* observing the direction of information flow and then backtracking the routing path up to the source node, finally killing the panda as shown in Fig. 1a. Fig. 1a shows that the source node (SN) (represented by the blue dot) sends the location of the asset (panda in this case) to the BS using SPR, shown with black colored arrows. The hunter, who is at the BS, receives the packet and moves in the direction of the packet's arrival and towards the SN one hop at a time as shown with red colored dotted arrows. For every new packet sent by the source, the attacker moves one hop toward the SN. The backtracking happens in time equivalent to the number of hops between the SN and BS.

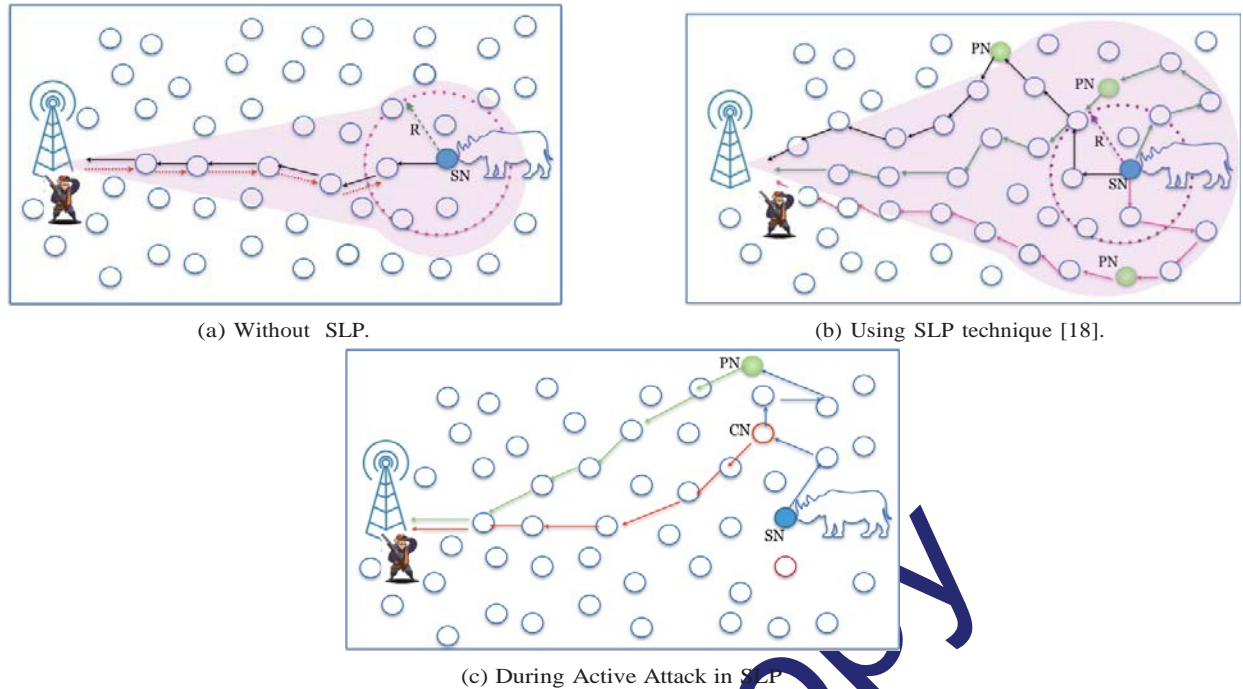

(a) Without SLP.

(b) Using SLP technique [18].

(c) During Active Attack in SLP

Fig. 1: Illustration of asset protection in different scenarios

In this context, Source Location Privacy (SLP) protection techniques have been devised to mitigate these issues. Such techniques increase the adversity of the attacker by either delaying the packet transmission over a longer randomized route or by creating fake SNs. It has been shown in the literature that such techniques can prove very effective in foiling the passive efforts of an attacker. SLP-aware routing technique was first introduced in the pioneering work in [18]. In this technique, instead of sending the packets to the BS via the shortest path, the source node employs a random selection of the neighbor and relays the packet to that node. The process of random neighbor selection continues till the design parameter $time\text{-}to\text{-}live$ (TTL), which is measured in terms of the number of hops, remains non-zero. Once the random walk is completed, the pack-ets are sent to the BS using the SPR technique. This is shown in Fig. 1b, where the SN (represented as a solid blue dot) is sending three packets; one via the black-colored path, another one via the blue-colored path, and the last one via the red-colored path. In the literature, the node where the random walk terminates is termed as $phantom\ node$ (PN), shown in green-colored solid circles in Fig. 1b. The objective of such approaches is to increase the routing path diversity that in turn increases the attacker's $backtracking\ period$. The effect of routing path diversity is pictorially demonstrated by the increase in the pink shaded area in Fig. 1b compared to that in Fig. 1a. An increase in $backtracking\ period$ in turn leads to an increase in the $safety\ period$ (See Section 5) which is one of the measures of privacy protection.

In this context, the literature is rich with SLP solutions to counter the efforts of a passive attacker. However, to the best of our knowledge, there are no works that explore the impact of active attacks on the performance of the said SLP solutions which is the premise of the current work. For instance, consider that the attacker has compromised a few nodes in the network as shown in red-colored circles in Fig. 1c. Let us assume that the packet sent by the source node reaches the compromised node (CN) after traveling for two hops and the initial $TTL$ value was set to 4 by the SN. On reaching the CN (represented by a red circle in Fig. 1c) the *TTL* value is reduced to 2 from 4 since the packet has traversed through 2 hops in reaching the concerned node. At this point, the CN can perform an active attack by abruptly changing the *TTL* value to zero and ending the random routing. The CN then relays the packet to the next node that is situated towards the sink, thus reducing the length of the random walk and the *backtracking period* of the attacker which in turn lowers the *safety period*.

Thus in this work we intend to study the effectiveness of TTL-based random walk SLP techniques when faced with an active attack. In this context the contributions of the present article can be summarized as follows:

– Discuss the motivation behind *active attacks* in WSNs and propose a hybrid attacker model consisting of both active and passive attacks;
– Present a system model that explains the implementation of the proposed hybrid model;
– Assess the impact of the proposed hybrid attacker model on the privacy performance of three existing TTL-based random walk SLP techniques.

The rest of the article is organized as follows: Section 2 describes the existing attacker models under various categories based on the capabilities of the attacker(s) considered in the literature; followed by a summary of the random walk-based SLP solutions available in the literature in Section 3. The three TTL-based random walk algorithms used as references in this work are also briefly outlined in this section. Section 4 presents relevance of the proposed hybrid attacker model and the system model supporting it. The experimental results and discussions on the research findings are presented in Section 5. Finally, the article ends with conclusions and future research directions as detailed in Section 6.

## 2 Existing Attacker Models

The attack models considered in all the existing works related to SLP in WSNs can be broadly categorized into three classes based on the adversary's capabilities: i) local attacker model [18], ii) global attacker model [15] and ii) semi-global attacker model [19]. We briefly present these models in the current section.

### 2.1 Local Attacker Model

Under this model, the adversary is considered to have only the local view of the network (i.e., one hop or one node visibility). It is assumed that the adversary commences its attack journey from the BS location as this information is open to everyone, including the adversary. The adversary chooses this location as the starting point because the entire network traffic converges at this location. Then, for every new packet that is arriving towards the sink, the adversary estimates the direction of the signal arrival, say location prediction techniques, and performs signal (or packet) correlation analysis. Based on the inferences made it moves ahead by one hop in the direction of the packet arrival. As the packets keep arriving at the sink, the adversary keeps moving one-hop toward the source of information. This process is repeated until it reaches the source of information. Finally, the source location is disclosed. This attack process is shown in red-colored dotted lines in Fig. 1a.

### 2.2 Global Attacker Model

The global attacker models were first seen in the works of Kiran Mehta et al. [16], and Min Saho et al., [26]. The adversary is assumed to be highly motivated and resource-rich. Particularly, under this model, the adversary deploys several snooping devices in the network. The number of such devices is almost equal to the number of sensor nodes in the target network. The job of devices is only to overhear the communications among the sensor nodes and pass this information to the adversary. However, these snooping devices cannot detect the assets or objects in the network. Such an attacker is believed to pose a significant threat to the assets that are under observation. However, it is also argued that such attack models are unrealistic because of the size of the adversary's network.

### 2.3 Semi-Global Attacker Model

In this model, the attacker is assumed to monitor the signal flow for more than one hop but not the entire network [19]. A variant of this model is multiple local (one-hop view) adversaries. The adversaries spread themselves throughout the network and collaborate by exchanging information on the traffic that they collect [9]. Then they reach the source of information based on the conclusions drawn from the attack analysis. In another variant, a single adversary distributes a few monitoring devices in areas of interest [11]. The adversary then uses these devices to collect the information and performs analyses of the data so that it can determine the location of the source.

The behavior of the attacker under these three models is classified either as a *passive attacker* or an *active attacker*. The *passive attacker* only eavesdrops on the communication links without affecting the normal network operations; whereas, the *active attacker* performs attacks such as packet dropping, packet content modification, injection of fake packets, compromise the nodes to gain access to secret information, or even block traffic at certain parts of the network. However, to the best of our knowledge, the majority of the SLP solutions try to address the *passive attacker* problem under the local-, global- and semi-global attacker models. It is believed that active attacks on WSN are easily determined using intrusion detection systems (IDS) and hence the attacker does not prefer such attacks. Nevertheless, as shown in [14] it is observed that conflicts do exist between IDS and SLP solutions developed for WSN. Therefore, given the advancement in technology and computational resources, we strongly believe that it is not prudent to assume that an attacker is only capable of conducting *passive attack*. It is thus necessary to explore the impact of *active attack* on the existing SLP solutions since an adversary can choose to use the same if it suits the need of the adversary. In the subsequent section, we initially describe the random walk-based SLP solutions existing in the literature followed by a brief outline of the three TTL random-walk-based SLP techniques considered to study the impact of *active attack*.

## 3 Random Walk-based SLP Solutions - A Summary

As mentioned in the earlier sections, there are several SLP solutions that are developed to address the *passive attack* problem. The solutions can be broadly classified as: i) Random path routing-based SLP techniques and ii) Fake packet-based SLP techniques. A detailed survey can be seen in [5]. However, since the focus of the current work is on studying the impact of *active attack* on the performance of the first category of solutions a brief summary of the existing solutions in the same is presented in this section. In the *random path routing-based* approach, the data packet from the source node is relayed to a neighbor node in a random fashion. This class of SLP can be further grouped into two classes: i) Biased random walk-based SLP schemes and ii) TTL-based random walk SLP schemes. In the former case, the packets are sent to the BS using the forward random walk (FRW) approach [3,4,18]. In this case, a node relays the information packets randomly to a neighbor that is lying toward the BS thus creating forward bias in the random walk. This approach continues till the packet reaches the BS. In the latter case, the packet initially is routed randomly (either in a pure random walk or in a biased random walk) till the TTL value, which is initialized by the SN, becomes zero. This approach is termed a phantom routing scheme (PRS) or phantom-based routing scheme (PBR). The TTL value is decremented by one after each random hop and once it becomes zero the packets are either directly sent to the BS using the shortest path routing or are routed either in the clock or anti-clockwise directions for another set of hops and then finally sent to the BS.

Among these two broad categories, it is shown that TTL-based random walk schemes does better in terms of safety period than the FRW-based routing schemes [20]. Therefore, in this work, we focus on only TTL random walk-based SLP schemes to study the impact of the proposed hybrid attacker model on some of the said schemes. The premise of these approaches is to entice the hunter or the adversary away from the source so that its backtracking period increases. However, it is observed that in case of pure random walk the deviation from the SN is limited [18] as a result of which, the backtracking time (measured in terms of hops) of the attacker might not significantly improve. To overcome this issue, a directed or biased random walk is often suggested for the random walk phase [13, 17, 18, 20]. These solutions aim to increase the randomness of the routing path taken by a packet from the SN which in turn leads to an increase in the *backtracking period* of the attacker.

After delineating the various random walk-based routing protocols available in the literature we focus on the three TTL random walk-based SLP protocols that serve as the basis of our current study in the next subsection.

3.1 TTL random walk-based routing protocols used as case studies

In order to assess the impact of the proposed hybrid attacker model we consider three existing TTL random walk-based SLP techniques: i) Phantom routing scheme (PRS) also termed as phantom-based routing (PBR) scheme [18], ii) source location privacy using randomized routes (SLP-R) [20], and iii) position-independent section-based scheme (PSSLP) [17]. An overview of the three techniques is provided in this subsection.[1].

In PRS technique, the SN upon detection of the asset, generates packets, initializes them with the TTL, and randomly selects a neighbor from its neighbor list to relay the packet to that node. The neighbor node receives the packets, decrements the TTL value by one, and then picks up the next relay node from its neighbor list, and then relays the packet to that neighbor. This process continues till the TTL value in the packet is non-zero. When the TTL value is zero, the node that has this packet (termed a PN) chooses the next relay node that is closest to the BS from its neighbor list. Every node in the path follows a similar approach to relay the packet to the BS that constitutes the shortest path routing (SPR).

---

[1] We use either PRS or PBR interchangeably in this work

SLP-R technique as proposed in [20] has three phases: i) Backward random walk (BRW) phase, ii) Identical depth routing (IDR) phase, and iii) Min-Hop Routing (MHR) phase. In the first phase the packet from the SN is sent away from the BS for a certain number of hops till the TTL value goes to zero. The node that receives the packet at the end of this phase is known as PN. In the next phase (IDR), the packets are routed using the nodes that are at a similar distance as the PN from the BS either in a clock or anti-clockwise direction which is randomly chosen for a predetermined number of hops. In the final phase the packets are sent to the BS using MHR. Overall, this gives the effect that the information packets are arriving at the BS from all directions in the network thereby increasing the routing diversity which in turn adds to the confusion of the attacker.

To eliminate the source node position-dependent behavior of the SLP protocol such as PBR and SLP-R, the authors in [17] proposed an improved SLP scheme named PSSLP. In this technique, nodes in the network are grouped into three segments (each of size one-third the network radius). Based on the segment in which the SN is present, the routing protocol either uses a TTL-based BRW, a simple BRW till the network edge, or no random walk. Further details about the algorithm can be found in [17].

In the next section, the relevance and description of the proposed hybrid attacker model followed by the system model supporting it is presented.

## 4 Proposed Hybrid Attacker Model

In this section, we first briefly explain the premise of a TTL-based random walk algorithm and conclude the section with a description of the proposed hybrid attacker model for such kinds of algorithms.

### 4.1 Relevance and Description

As described earlier any TTL-based random walk protocol comprises primarily two phases - TTL-based random walk and shortest distance routing. In the first phase, the SN initializes a non-zero TTL value to decide the number of hops for the random walk propagation. After each hop, the TTL value is decremented by one by the current node holding the information. This phase might have other additional strategies to increase the randomness of transmission. The node reached at the end of this phase is known as the PN. Once the PN is reached then transmission to the BS happens through SPR. Fig. 1b depicts one such algorithm where the TTL value is initialized to 4.

The primary objective of a passive attacker is to backtrack the transmission direction using the direction of arrival of data packets to identify the location of an SN. In this context, the previously mentioned TTL-based random walk protocols increase the backtracking period by randomizing the data propagation as per some strategy that is not known to the attacker. However, in order to negate the impact of randomized routing the attacker can adopt a hybrid approach comprising of active and passive attack. Initially, for the implementation, the attacker backtracks for a few hops to estimate the direction of arrival of data. Once the direction of arrival is estimated the attacker uses an active attack. In the active attack mode, the attacker compromises nodes mostly in the estimated direction of arrival. To cite an example, let us assume that the attacker analyses through a few hops of backtracking that data packets arrive from a direction to the right of the BS and decide to compromise a total of $C$ nodes in the network. In this scenario, the majority of the $C$ nodes will be to the right of the BS while a few will be to the left. If a CN falls in the TTL-based random walk path then the CN reduces the TTL value to zero irrespective of its current value. The CN then behaves as the PN and initiates the shortest path routing to the BS. In case the CN does not fall in the random walk path it behaves like a normal node and does not affect data routing. This approach of the attacker weakens the privacy strength of the SLP protocols. The corresponding system model is presented in the next section.

### 4.2 System Model

We model the network of sensor nodes as a network graph consisting of $N$ number of sensor nodes and $E$ number of links among the nodes similar to [2]. The graph $G$ is a pair represented as $G = \{N, E\}$, where $N = \{n_0, n_2, n_2, \ldots n_{N-1}\}$ and $E = \{e_1, e_2, e_3, \ldots e_M\}$, $M$ is the total number of wireless links in the network. Further, the neighbor list of a sensor node $n_i$, with $i \in N$, is represented by $\Phi_{n_i} = \{n_1, n_2, n_3 \ldots n_K\}$ where $K$ is the number of neighbors of node $n_i$, i.e., $K = |\Phi_{n_i}|$. Let $d(n_i)$ denote the shortest distance between the node $n_i$ and the base station $n_0$ measured in hops, then obviously $d(n_0) = 0$. $R = [r]_{N \times N}$ denotes the transition matrix. An element $r_{ij}$ in the matrix represents the transition probability between nodes $n_i$ and $n_j$. If there is no link between $n_i$ and $n_j$ then $r_{ij} = 0$. Thus there are $M$ non-zero elements in matrix $R$.

The routing process consists of selecting a neighbor as per the privacy-preserving algorithm and relaying the packet to its neighbor. This process continues till the packet reaches the BS. For this routing purpose, we define $\phi_{n_i}$ as the next-best neighbor of node $n_i$. If SPR is employed then $[R]_{[i,j]} = 1$ if $n_j$ is the next-best relay node of $n_i$ i.e., $d(n_j) = d(\phi_{n_i}) = \min_{n_k \in \Phi_{n_i}} d(n_k)$. In the TTL-based random walk phase the next best neighbor is selected as per the algorithm under investigation with the TTL value, $T$ used as one of the algorithmic parameters. In the current work, three different algorithms namely, PRS, SLP-R, and PSSLP are investigated.

In the case of both passive and hybrid attacks, the attacker starts from the BS and identifies the node $n_k$ from which data is received at the BS. It identifies the sequence of nodes through which data is routed by backtracking till it reaches the SN in case of a passive attack. In case of a hybrid attack, it uses the backtracking information to estimate the direction of arrival of data packets as right ($R$) or left ($L$) of the BS. The attacker compromises $N_C$ nodes in the network depending on the available resources, where $N_C \subset N$. If the estimated direction of arrival is $R$ then the majority of $N_C$ is to the right and is randomly chosen. Let $n_s$ denote the SN and $\Phi_{n_s} = \{n_s, n_{s-1}, \ldots, n_{s-m}\}$ denote the nodes in the TTL random walk path of data routing. Then $T = |\Phi_{n_s}| = m+1$, where $m+1$ is the number of nodes through which the data traverses in the TTL-based random walk phase. $T$ is reduced by one at every hop during this phase and the phase ends once $T = 0$ during normal operation. However, as per the active phase of the hybrid model if a CN, $n_c$ where $n_c \in N_C$ also belongs to $\Phi_{n_s}$ then $T$ is abruptly changed to zero irrespective of the current value. $n_c$ then behaves as the PN and initiates the SPR to the BS. This strategy of the attacker cuts short the TTL-based random walk phase and reduces the randomization of the data routing. But in the proposed model if $n_c \notin \Phi_{n_s}$ then it does not have any impact on the data routing.

In the next section, we describe the simulation environment and network settings used to assess the performance of the three reference TTL-based random walk solutions considered in this work. The impact of the proposed hybrid attack in comparison to the passive attack in terms of privacy is also presented in the context of the three algorithms.

## 5 Experimental Results and Discussion

In this section, we present the impact of the proposed hybrid attack model on the effectiveness of three existing TTL random walk-based SLP solutions. As alluded in Sec. 3.1, we use PRS, SLP-R, and PSSLP schemes as reference to assess the impact of the proposed attacker model in terms of privacy metrics such as safety period, capture ratio, and entropy. These metrics are defined as follows:

– *Safety Period*: It is defined as the number of packets sent to the BS before the attacker locates the SN [18].
– *Capture Ratio*: It is defined as the percentage of the total trials in which the attacker finds the location of the SN [18].
– *Entropy*: Entropy is the average amount of uncertainty in a given dataset. In the current context it is a measure of the randomness of the routing path. [6].

Since the prime focus of the work is to study the impact of the proposed attacker model on preserving the privacy of SN using existing SLP techniques, the performance of the techniques is studied only in terms of the privacy metrics described earlier. The simulation scenario used for the comparison is presented next.

### 5.1 Simulation Scenario

The simulations were carried out using Python 3.9.0. We consider a square deployment of sensor nodes over an area of 2 km × 2 km. The sink also known as the BS is located at the center of the generated network with coordinates of (0, 0). The spacing (which we term as offset value) between each sensor is set to 50 m (using the condition $r_s/\sqrt{2}$, where $r_s$ represents the radio range of a sensor node). Based on this offset value, the radio range of the sensor nodes $r_s$ turns out to be 72 m. With this configuration, there would be 1600 sensor nodes in the network and the network density (which is the ratio of the total number of sensor nodes to the total area of the network) is approximately 0.8. During simulation, 500 packets are sent from the SN per trial and a total of 50 trials were performed per position of the SN. Three different positions of the SN at distances of 300, 600, and 900 m from the BS were considered for each of the scenarios for any of the explored SLP techniques. The size of the data packet transmitted was kept constant at 2048 bits. The simulation in each trial ends when the attacker finds the location of the SN. While the value of TTL was set to 5 for all the reference techniques considered, the value of hop count in phase two of SLP-R was randomly chosen between 0° and 180° respectively, in terms of number of hops [20]. The simulation parameters are the same as those considered in [13, 17, 20].

In order to compare the privacy effectiveness of the considered techniques in the face of the proposed hybrid attacker model in contrast to a passive attacker model two scenarios were considered in the simulations,

1. Scenario I: No CN which as per our current description of the hybrid attacker model is equivalent to saying that the attacker is passive;
2. Scenario II: Non-zero number of CNs which as per our presented model represents a hybrid strategy of the adversary consisting of both active and passive attacks.

In Scenario II, to assess the impact of the number of CNs on the privacy effectiveness of the reference SLP techniques, the number of CNs were varied in the range 5 to 20 in steps of 5 for each technique and privacy metric. We present the impact of the hybrid attacker model in contrast to a passive attack on the performance metrics of the three reference SLP schemes, in the next section.

5.2 Effect of hybrid attacker model on the performance of PRS, SLP-R, and PSSLP techniques

In this subsection we sequentially present the impact of the hybrid attacker model on the performance of PRS, SLP-R, and PSSLP algorithms and contrast it with their corresponding performance during a passive attack in terms of different privacy metrics.

5.2.1 Performance of PRS

Fig. 2a presents the performance of the PRS algorithm in the face of a hybrid and a passive attack in terms of $safety\ period$. While the black solid line corresponds to Scenario I the other lines correspond to Scenario II with varying degrees of hybrid attack. The plots clearly show that the impact of the hybrid attack on the PRS technique in terms of $safety\ period$ is more pronounced compared to a passive attack irrespective of the number of CNs and position of the SN. Another interesting observation in this context is the result with 10 CNs where the performance is the worst especially when the SN is closer to the BS rather than the edge of the network. This shows that the adversary will be at an advantage if it compromises only an optimum number of nodes rather than mindlessly compromising as many nodes as possible.

The performance of PRS technique in terms of $capture\ ratio$ is shown in Fig. 2b. As before the black solid line indicates the performance of the technique when the attacker performs a passive attack. As indicated by the plots, it can be observed that the attacker has a better chance of finding the location of the SN if it performs a hybrid attack as opposed to a passive attack. Furthermore, similar to the result with respect to $safety\ period$ the performance of the technique is most affected by the hybrid attack when the number of CNs is 10. The result reaffirms that the usage of an optimum number of CNs can be more advantageous to the adversary.

It can be clearly observed from Fig. 2c that the performance of the PRS technique is better in Scenario I when a passive attack is launched even in terms of $entropy$ compared to a hybrid attack irrespective of the severity of the attack (i.e. number of CNs). It can also be noted that just like the other privacy metric, entropy is affected most when there are 10 CNs and the SN is placed either at a distance of 300 or 600 m from the BS. Thus it can be inferred that irrespective of the privacy metric compromising 10 nodes when the SN is closer to the BS compared to the edge of the network affects the performance of the PRS technique most for the investigated simulated environment. The results in regards to SLP-R is presented next.

5.2.2 Performance of SLP-R

Fig. 3 presents the performance of SLP-R in the face of passive (Scenario I - solid black lines) and hybrid (Scenario II - the remaining colored lines in the plots) attacks. It can be clearly observed from the obtained results that irrespective of the performance metric ($safety\ period$ in Fig. 3a, $capture\ ratio$ in Fig. 3b or $entropy$ in Fig. 3c) the protocol provides a more secure solution when the network is under passive attack. The privacy effectiveness reduces drastically when the attacker adopts a hybrid approach as proposed in this work.

In this light, it is interesting to note that like the PRS algorithm, the performance degradation of SLP-R is maximum when the number of CNs is 10 and the SN is at a distance of 600 m from the BS. In addition, it can also be observed that when there are 20 CNs the performance of the algorithm in terms of privacy effectiveness is the best at the previously mentioned location of the SN. As a final observation, it can be noted that for each location of the SN, the impact of the number of CNs is not similar. To cite an example, when the SN is at 300 m from the BS 5 CNs have the prominent effect, 15 CNs are optimum from the attacker perspective when the SN is at 600 m from BS whereas comprising 5 or 20 nodes affects the privacy effectiveness most when the SN is at the network edge (900 m from BS). The result re-affirms the observation made from the results with the PRS algorithm - for the given hybrid model there is an optimum number of CNs for a given source location in a network. The performance comparison of PSSLP in the face of passive and active attacks is presented in the next subsection.

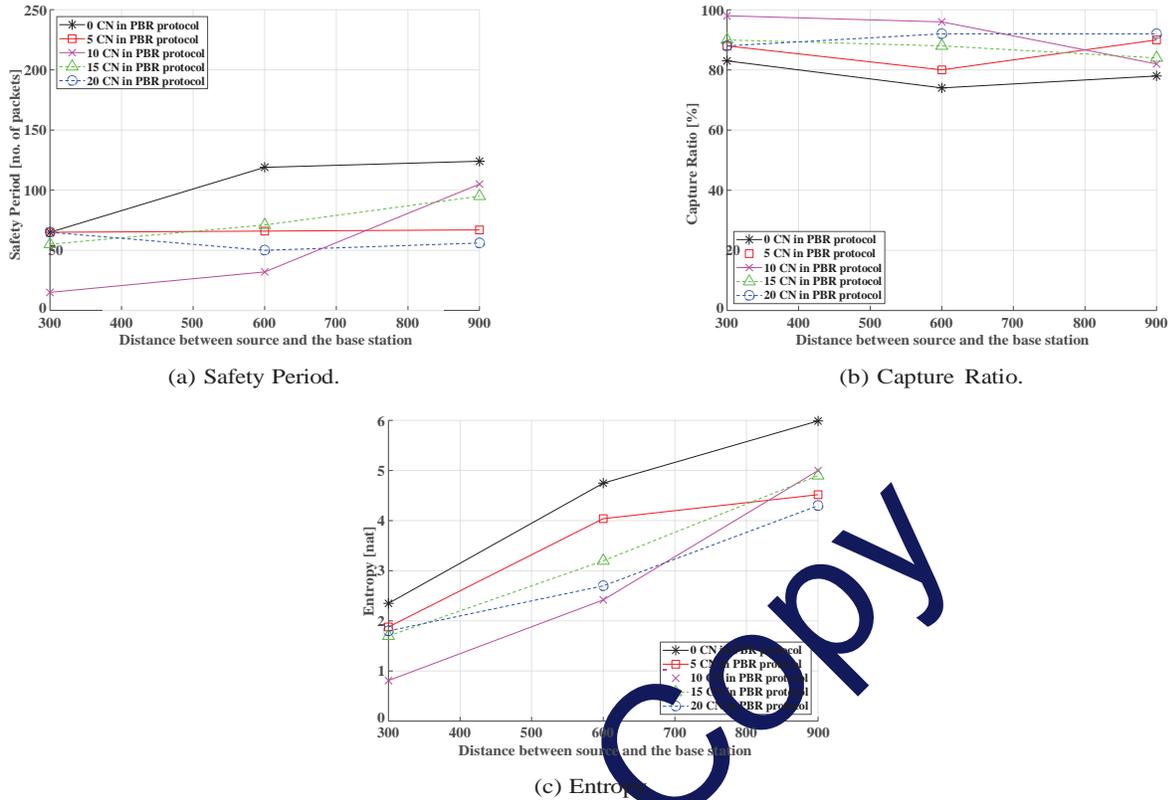

Fig. 2: Effect of hybrid attacker model on privacy effectiveness of PRS technique.

5.2.3 Performance of PSSLP

The performance of the different privacy metrics for the third protocol, PSSLP, are shown in Fig. 4. While Fig. 4a presents the performance in terms of *safety period*, Fig. 4b and 4c present the results in terms of *capture ratio* and *entropy* respectively. As before the plots in the black solid line correspond to passive attack (Scenario I) whereas the other plots are for hybrid attack (Scenario II) with varying numbers of CNs.

As for the results it can be clearly observed that the proposed hybrid attack negatively affects the performance of PSSLP in terms of privacy effectiveness for most of the SN locations. An exception in this regard is when the SN is placed at the edge of the network (900 m from the BS). In this scenario, it can be observed that if 15 or 20 nodes are compromised then the privacy performance of the algorithm is either equivalent in terms of certain parameters (*entropy*) or is better (*safety period* and *capture ratio*). In the context of the other source locations investigated in this work, privacy is severely affected when 10 nodes are compromised and least affected when the number of CNs is 20. In this regard, it can be noted that 10 CNs is an exception otherwise the negative impact of a hybrid attack on the performance of the algorithm from the privacy perspective compared to a passive attack reduces with increasing CNs. A summary of the results and insights are presented in the next subsection.

5.3 Discussion and Insights

Although it is generally assumed in source location privacy preservation problems that the attacker is passive, the results from the current investigation indicate that an active attack is possible and it can adversely affect privacy preservation. This necessitates the need for developing new SLP solutions that can address not only passive attacks but active attacks as well. One possible solution in this context could have been the IDs.

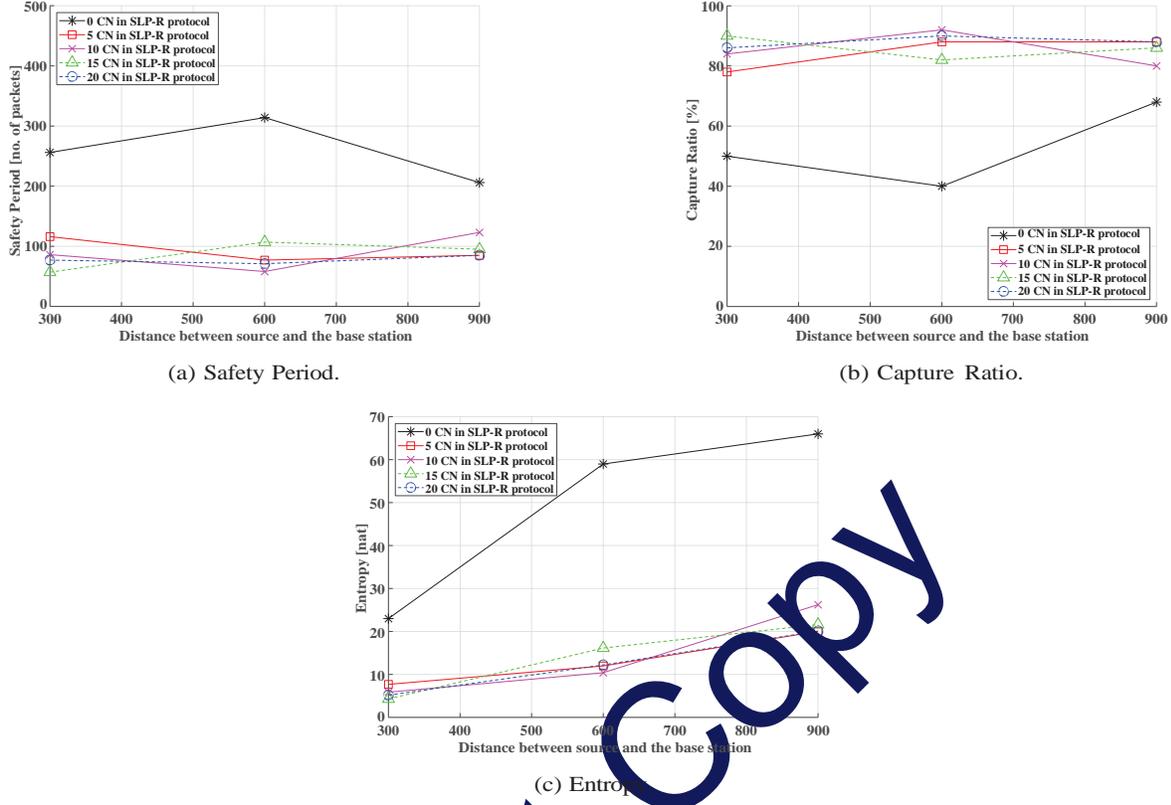

Fig. 3: Effect of hybrid attacker model on privacy effectiveness of SLP-R technique.

Table 1: Summary of Performance Metrics Degradation of PRS, SLP-R and PSSLP Techniques

| Protocol | Capture Ratio | Safety Period | Entropy |
|---|---|---|---|
| PRS | 13.81 | -7.46 | -9.84 |
| SLP-R | 71.57 | -65.73 | -72.35 |
| PSSLP | 125 | -83.58 | -17.36 |

However, the authors in [14] have shown that providing privacy and intrusion detection are orthogonal tasks. An alternate approach can utilize the secret sharing solutions presented in [8, 10, 12, 22, 25] which has never been used in the context of source location secrecy.

Table 1 presents the percentage decrement in performance of the reference TTL-based random walk protocols used in this work. The table shows the average values of all the network settings, averaged over all trials and possible scenarios for each investigated technique. While the negative values in the table indicate a decrease in the corresponding metric, the positive values indicate an increase in the corresponding metric. In light of source location protection, however, both values indicate a degradation in performance. To cite an example, Table 1 shows that there is a 125% increase in *capture ratio* for PSSLP when the hybrid attack as proposed in this work is used instead of a passive attack. The number indicates that in the case of a hybrid attack, the attacker is 125% more prone to identify the source location compared to a passive attack. Similarly, in the case of PRS and SLP-R, the corresponding increase is 13.81 and 71.57 percent respectively. In terms of *safety period* and *entropy*, the negative numbers indicate the percentage decrease in privacy protection in regard to the respective metric when the hybrid attack model is employed by the attacker. However, the numbers from the table must be read with caution keeping in mind the fact that irrespective of the mode of attack (passive or hybrid), PSSLP provides maximum protection on average as indicated by the values of each metric followed by SLP-R and PRS. Thus, although the average degradation in performance in terms of privacy metric is maximum in PSSLP when there is a hybrid attack, it can be observed from Fig. 4 that for certain cases (SN near the edge and number of CNs is 15 or 20)

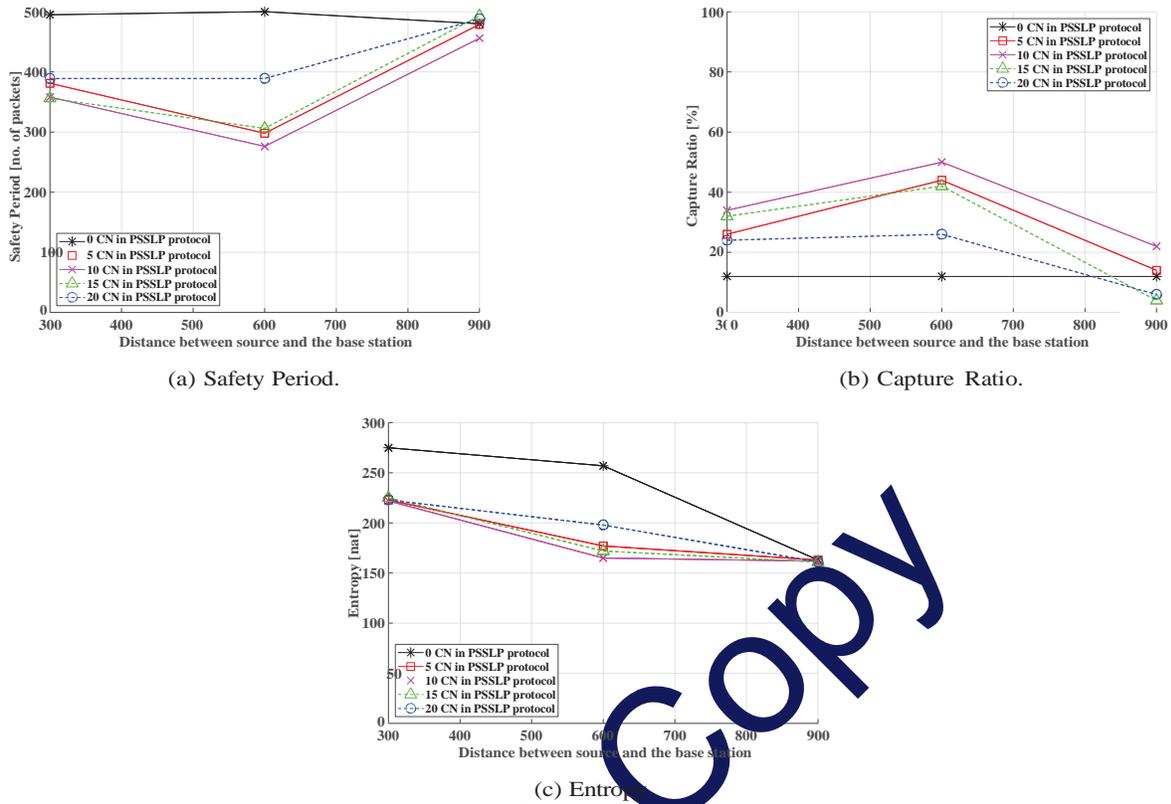

Fig. 4: Effect of hybrid attacker model on privacy effectiveness of PSSLP technique.

the performance actually improves. This is an interesting revelation.

Another observation we made is the choice of the right number of CNs' to positively favor the attacker. Our findings indicate that in the considered network setting the performance degradation is generally maximum when 10 CNs are used compared to 5, 15, or 20 CNs. However, this observation cannot be generalized and is possibly network as well as protocol-dependent. For instance, the table has clearly shown that PSSLP is on an average most affected by hop count modification attacks compared to PRS and SLP-R. However, as explained and observed earlier from the results, PSSLP performance actually improves in certain cases when there is a hybrid attack. The reason for such varied performances can be attributed to

- the variation in the amount of randomness introduced by a protocol;
- the number of CNs and their positions in the network.

## 6 Conclusion

In this work, we suggested a hybrid attacker model for WSNs as applied for event monitoring scenarios. In particular, we emphasized the impact of hop count modification attacks by the CNs on existing TTL-based random walk protocols. Through simulations, we have demonstrated that hybrid attacks i.e., TTL modification along with passive attacks can significantly degrade the performance of existing source location privacy protection schemes. We tested the impact of the introduced hybrid attacker model on three existing SLP protection techniques namely, PRS, SLP-R, and PSSLP. It was observed that although the PSSLP protocol was most affected on average by the proposed attack, there were certain scenarios where the privacy protection performance of the algorithm actually improved. Such a unique result was not observed for the other reference algorithms. A further noteworthy observation was in regard to the optimum number of CNs and their relative position to the SN which degraded the performance of the protocols the most. As a future extension to the current work, the hybrid attacker model can be fur-ther generalized such that impact is not limited to only TTL-based random walk solutions. In addition, a solution to counter the impact of such a generalized hybrid attack model can be developed.

**Declarations**

**Competing Interests:** No competing interests.
**Compliance with Ethical Standards:** The authors assure adherence to rules and ethical policy and declare that the work does not contain any human or animal involvement.
**Research Data Policy and Data Availability Statements:** The work does not contain any data set that could be shared. Furthermore, no other data or code is shared or made available online as the research is still in progress.